\documentclass[aps,prl,twocolumn,amsmath,amssymb,floatfix]{revtex4}
\usepackage{tabularx}
\usepackage{bm}
\usepackage{euscript}
\usepackage{epsfig,psfrag,subfigure}
\usepackage{graphicx}
\usepackage{color}
\usepackage{amsfonts}
\usepackage{exscale}
\usepackage{amsbsy}
\begin{document}
\title{An exact chiral spin liquid with non-Abelian anyons}
\author{Hong Yao and Steven A. Kivelson}
\affiliation{Department of Physics, Stanford University, Stanford, CA 94305}
\date{\today}
\newcommand{\br}{\mathbf{r}}
\newcommand{\be}{\mathbf{e}}
\newcommand{\bk}{\mathbf{k}}
\newcommand{\bK}{\mathbf{K}}
\newcommand{\bp}{\mathbf{p}}
\newcommand{\bq}{\mathbf{q}}
\newcommand{\bR}{\mathbf{R}}
\newcommand{\bS}{\mathbf{S}}
\newcommand{\bs}{\mathbf{s}}
\newcommand{\bb}{\mathbf{b}}
\newcommand{\bx}{\mathbf{x}}
\newcommand{\by}{\mathbf{y}}
\newcommand{\te}{\mathrm{e}}
\newcommand{\eff}{\mathrm{eff}}
\newcommand{\sgn}{\mathrm{sgn}}
\newcommand{\Det}{\mathrm{Det}}
\newcommand{\Tc}{\mathrm{T}_\mathrm{c}}
\begin{abstract}
We establish the existence of a chiral spin liquid (CSL) as the {\it exact} ground state of the Kitaev model on a decorated honeycomb lattice, which is obtained by replacing each site in the familiar honeycomb lattice with a triangle. The CSL state {\it spontaneously} breaks time reversal symmetry but preserves other symmetries. There are two topologically distinct CSLs separated by a quantum critical point. Interestingly, vortex excitations in the topologically nontrivial (Chern number $\pm 1$) CSL obey {\it non-Abelian} statistics. 
\end{abstract}
\maketitle
There has been enormous interest recently in the notion of spin-liquid states with fractionalized quasiparticle excitations, especially those that exhibit fractional statistics.  Since, to date, unambiguous experimental evidence of fractionalization has only been achieved in the fractional quantum Hall effect (FQHE), it is reasonable to seek a high standard of theoretical control in making statements about the behavior of model systems that are being studied in this context.  A breakthrough occurred when Moessner and Sondhi~\cite{Moessner01} demonstrated the existence of a symmetry preserving spin-liquid phase in the quantum dimer model~\cite{Rokhsar88}, analogous to  the short-range version~\cite{Kivelson87} of the RVB state proposed~\cite{Anderson87} to be the key to understanding high temperature superconductivity.  This state has fractionalized excitations, but only with Fermi or Bose statistics. Motivated by a  formal analogy between frustrated interactions and a magnetic field, Kalmeyer and Laughlin (KL)~\cite{Kalmeyer87} proposed a variational spin-liquid state for the triangular lattice quantum antiferromagnet (AF) constructed in analogy with the Laughlin state~\cite{Laughlin83} of the FQHE.  This state breaks time reversal symmetry (TRS)~\cite{Kivelson88}, and is thus a first example of a proposed class of chiral spin liquids (CSL)~\cite{Wen89}. However, both analytical and numerical studies~\cite{Bernu94} suggest that the AF Heisenberg model on a triangular lattice possesses $\sqrt{3}\times\sqrt{3}$ Neel order, which breaks translational and spin rotational symmetries in addition to TRS. Therefore, it remains conceptually interesting to determine whether a 
CSL could exist as a stable ground state phase of any Hamiltonian with only short-range interactions. 

\begin{figure}[b]
\subfigure[]{
\includegraphics[scale=0.16]{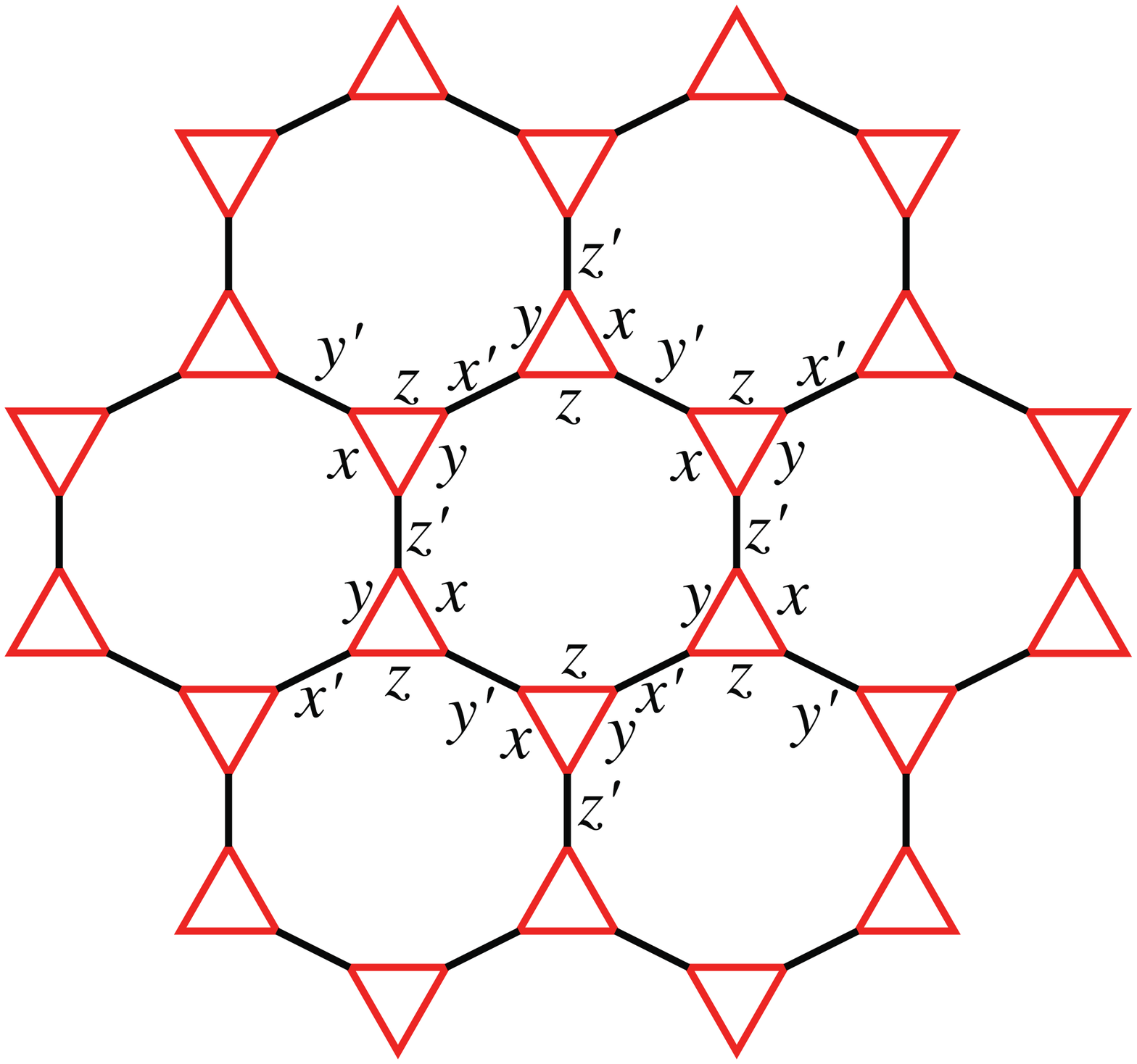}}
\subfigure[]{
\includegraphics[scale=0.16]{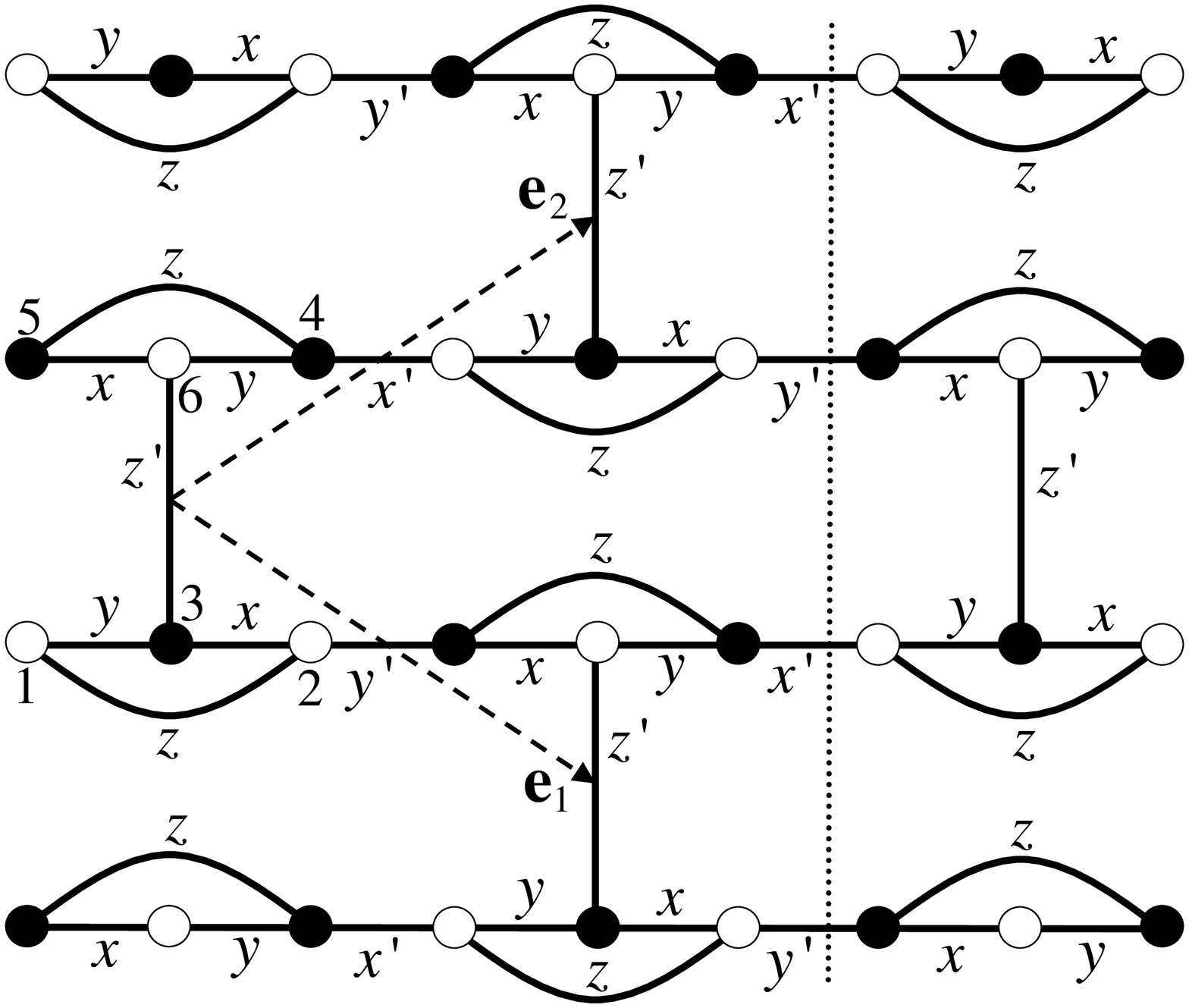}}
\caption{(a) The triangle-honeycomb lattice is constructed from a honeycomb lattice by replacing each site with a triangle. 
(b) Topologically equivalent representation. Sites within a unit cell $\br$ are labeled by $\alpha=1,\cdots,6$. 
The dotted line indicates its vertical boundary. }\label{fig_triangle_honeycomb_lattice}
\end{figure}

Here, we study the Kitaev model~\cite{Kitaev06} on the  ``triangle-honeycomb lattice'' (Fig.~\ref{fig_triangle_honeycomb_lattice}(a)) and show that its exact ground state is a CSL -- {\it i.e.} a state which spontaneously breaks time-reversal symmetry but neither spin rotational nor translational symmetry, and which has fractionalized excitations.   
Indeed, there are two topologically distinct CSL phases (with even and odd Chern number, respectively), separated by a quantum phase transition. For the topologically non-trivial (trivial) CSL, the vortex excitations obey non-Abelian (Abelian) statistics. (There are, however, an even number of electrons per unit cell, so this state may not deserve to be called a ``Mott insulator''.)  As far as we know, this is the first model whose {\it exact} ground state is a CSL.
  
The ``triangle-honeycomb'' lattice shown in Fig.~\ref{fig_triangle_honeycomb_lattice}(a) is obtained by replacing each site of a honeycomb lattice with a triangle - there are thus 6 sites per unit cell.   
It is important to note that, in contrast to the   
honeycomb lattice, the elementary plaquettes of this lattice have an {\it odd} number of sites.  The Kitaev model (Eq.~(\ref{eq_model})) on this lattice is time reversal invariant. It is also invariant under inversion  
of the lattice and under $180^\circ$ rotation of the spins along the $\hat x$, $\hat y$ or $\hat z$-directions, {\it i.e.} spin-orbit coupling has reduced the spin symmetry from  $SU$(2) to $D_2$. 
We will show that the ground state of the Kitaev model on this lattice does not break the spin symmetries, nor any lattice translational or 
point group symmetries. (Indeed, the spin correlation function is exactly zero beyond nearest neighbors~\cite{Baskaran07}.)  
However, in both the topologically trivial and non-trivial states, the exact ground state {\it spontaneously} breaks TRS. 
Similar states were studied previously by Kitaev~\cite{Kitaev06} and Lee {\it et al}~\cite{Lee07} on a honeycomb lattice,
 but in models in which TRS is explicitly broken, {\it e.g.} by an external magnetic field. 

The existence of an exactly solvable model with a CSL ground-state is suggestive that such phases might exist in nature. One might try to realize something close to this precise model in a system of  cold atoms  
in an optical lattice~\cite{Duan03}.  More generally, the fact that quantum Hall physics and non-Abelian anyons can occur without the need for a large applied magnetic field, opens the possibility (still very distant, at present) that some of this interesting physics might be seen on higher energy scales and in simpler systems than was previously realized.

{\it Model Hamiltonian}: 
The Kitaev model on the triangle-honeycomb  lattice can be written as follows:
\begin{eqnarray}\label{eq_model}
&&{\cal H}=\sum_{x\textrm{-link}} J_x\sigma^x_i\sigma^x_j +\sum_{y\textrm{-link}}J_y\sigma^y_i\sigma^y_j +\sum_{z\textrm{-link}}J_z\sigma^z_i\sigma^z_j\nonumber\\ &&+\sum_{x'\textrm{-link}} J'_{x}\sigma^x_i\sigma^x_j +\sum_{y'\textrm{-link}}J'_{y}\sigma^y_i\sigma^y_j +\sum_{z'\textrm{-link}}J'_{z}\sigma^z_i\sigma^z_j,
\end{eqnarray}
where $\sigma^\alpha_i$ ($\alpha=x,y,z$) is a Pauli matrix on site $i$. In the summation, $i$ and $j$  are nearest neighboring sites connected by a $\alpha$-link. (See Fig.~\ref{fig_triangle_honeycomb_lattice}.)  
We assume that $J_\alpha$ and $J'_{\alpha}$ are positive in this paper
~\cite{comment_J_sign}. 
This model on a honeycomb lattice was studied originally by Kitaev~\cite{Kitaev06}. Since then, there have been some interesting developments~\cite{Xiang07,Baskaran07,Chen07,Lee07,Chen07b}. As shown by Kitaev, this model can be mapped to a free Majorana model with conserved background $Z_2$ gauge fields in an enlarged Hilbert space. The physical states are obtained by implementing a set of gauge constraints. Later, Feng, Zhang and Xiang~\cite{Xiang07} proposed a different method of solution, which we will adopt, based on transforming spins into Jordan-Wigner fermions.    

In order to implement a Jordan-Wigner transformation~\cite{Fradkin80}, we need to order the sites.    
In Fig.~\ref{fig_triangle_honeycomb_lattice}(b), the sites can  
be labeled by a chain index, $m$, and a site index, $l$ along the chain.  For $i=(m,l)$ and $j=(m^\prime,l^\prime)$, we define an ordering such that $i< j$ if $m < m^\prime$ or if $m=m^\prime$ and $l < l^\prime$.  For open boundary conditions (OBC) in the $\hat x$ direction, the origin of $l$ for this ordering is defined as the left edge of the system, while for periodic boundary conditions (PBC), we define the origin along an (arbitrarily defined) ``cut,'' as indicated by the dotted line in Fig. \ref{fig_triangle_honeycomb_lattice}(b). 
We can define the usual Jordan-Wigner fermions: 
\begin{eqnarray}
&&\sigma^x_{j}=(a^\dag_{j}+a_{j})\exp\big[i\pi\sum_{i<j}a^\dag_{i} a_{i}\big],\\
&&\sigma^z_{j}=(-1)^j(2a^\dag_{j} a_{j}-1),
\end{eqnarray}
where $a^\dag_j$ is a fermion creation operator and satisfies the usual 
anticommutation relations and $(-1)^j=1$ (-1) for black (white) sites.   
A notable feature of the Kitaev model is that, despite being two dimensional, because the only inter-row interactions occur on $z'$-links, the Hamiltonian expressed in terms of fermionic variables is still local.

A further simplification occurs when we decompose each complex fermion into two Majarona fermions $c_j$ and $d_j$, 
as follows: $c_j\equiv-i(a^\dag_j-a_j)$ and $d_j\equiv a^\dag_j+a_j$ on black sites while $d_j\equiv-i(a^\dag_j-a_j)$ and $c_j\equiv a^\dag_j+a_j$ on white sites. 
${\cal H}$ can thus be rewritten as 
\begin{eqnarray}\label{eq_Ham_gauge}
&&{\cal H}=\sum_{x,y\textrm{-link}}iJ_\alpha c_i c_j +\sum_{z\textrm{-link}} iJ_z\hat U_{ij} c_i c_j\nonumber\\
&&\qquad +\sum_{x',y'\textrm{-link}}iJ'_\alpha c_i c_j +\sum_{z'\textrm{-link}}iJ'_z\hat U_{ij}c_i c_j,
\end{eqnarray}
where  $\hat U_{ij}\equiv -id_i d_j$ is a set of Hermetian operators defined for all pairs of sites $ij$ connected by either a $z$-link or a $z^\prime$-link. An ordering convention is adopted so that the products $c_ic_j$ and $d_id_j$ are always taken with $i < j$ for all links. While Eq.~(\ref{eq_Ham_gauge}) is exact for OBC, for PBC an {\it additional} boundary term must be added:  ${\cal H} \to {\cal H} + \sum_{\textrm{cut}}iJ'_\alpha \hat F_m c_i c_j$  where 
the sum runs over links across the cut, 
and $\hat F_m=\exp[i\pi\sum_{k\in m}a^\dag_ka_k]=\prod_{k\in m} c_kd_k$ with $k$ summed over all sites on chain $m$ that contains link $ij$. This boundary term is crucial to obtain the correct topological ground state degeneracy and seems not to have been treated in previous studies. The operators $\hat F_m$ commute with each other and with ${\cal H}$, and so are constants of the motion with eigenvalues $F_m=\pm 1$. 

Manifestly, $\hat U_{ij}$ is idempotent, and can be thought of as a background $Z_2$ gauge connection. The set of $\hat U_{ij}$ commute with each other, and for OBC, with the Hamiltonian, as well. Thus, we can label distinct sectors of Hilbert space by the set of eigenvalues $\{U_{ij}\}$. The Hamiltonian operator in  this sector is ${\cal H}(\{U_{ij}\})$;  it is precisely of the same form as in Eq.~(\ref{eq_Ham_gauge}), but with the operators $\hat U$ replaced their eigenvalues, $\hat U_{ij} \to U_{ij} = \pm 1$. This reduces ${\cal H}(\{U_{ij}\})$ to a quadratic form!   

For PBC which we will focus on, the situation is somewhat more subtle since $\hat U_{ij}$ on any $z'$-link anticommutes with $\hat F_m$ for the chain $m$ containing $i$ or $j$, and consequently does {\it not} commute with ${\cal H}$.  In this case, we instead label the sectors of Hilbert space by a set of $Z_2$ fluxes, $\hat\phi_p=\prod_{ij}^p \hat U_{ij}$ where $p$ labels the elementary plaquettes (triangles and dodecagons) and the product is over the $z$ and $z^\prime$ links surrounding the plaquette. (For plaquettes straddling  the cut, two corresponding $\hat F_m$'s should be included in the product, as well.) For triangle plaquettes, $\hat \phi_p =\hat U_{ij}$. 
Since $[{\cal H},\hat \phi_p]=[\hat \phi_p,\hat \phi_{p'}]=0$,  the sectors of Hilbert space can be uniquely specified by the eigenvalues of the local flux operators, $\phi_p=\pm 1$, and by two global $Z_2$ fluxes, $\Phi_x$ and $\Phi_y$,  computed along an arbitrarily chosen contour encircling the system in the $\hat x$ and $\hat y$ directions, respectively.  The Hamiltonian in each sector, ${\cal H}(\vec \Phi,\{\phi_p\})$, is again a quadratic form, 
 obtained by replacing $\hat U_{ij}$ and $\hat F_m$ in ${\cal H}$ by $\pm 1$ consistent with the specified flux pattern $(\vec \Phi,\{\phi_p\})$ with the understanding that the true eigenstates of ${\cal H}$ should be obtained by applying an appropriate projection operator on the eigenstates of the quadratic Hamiltonian~\cite{comment_wave_function}.  The corresponding ground state energy is $E_0(\vec \Phi,\{\phi_p\})$.

{\it Broken time reversal symmetry}:
Under time reversal $\hat T$, $\hat U_{ij} \to -\hat U_{ij}$ on any $z$-link and $z'$-link; $\hat F_m\to \hat F_m $ for any $m$. Thus $\hat \phi_p$ is even on dodecagonal and {\it odd} on triangular plaquettes. The true ground-state is found by identifying the sector or sectors in which $E_0$ is minimal. In all cases, the ground-state of this model {\it must} break TRS, and must therefore be at least two-fold degenerate~\cite{comment_T_breaking}. To see this, suppose 
a set $(\vec \Phi,\{\phi_p\})$ which minimizes $E_0$. There must exist a distinct time-reversed set ($\vec \Phi^T,\{\phi^T_{p}\}$) such that $E_0(\vec \Phi^T,\{\phi^T_{p}\})=E_0(\vec \Phi,\{\phi_p\})$ where $\vec \Phi^T=\vec \Phi$ and $\phi_p^T=\phi_p$ ( $\phi_p^T=-\phi_p$) on dodecagonal (triangular) plaquettes. This twofold degeneracy due to TRS is in addition to the topological degeneracy discussed below.

{\it Uniform fluxes}: 
To find the global ground-state or ground-states, we need to identify the sector in which $E_0$ is minimal. For generic $J_\alpha$ and $J'_\alpha$, we have not yet obtained an analytic solution to this problem. 
However, we have analyzed~\cite{long} the problem using degenerate perturbation theory, both in the limit $J_\alpha\ll J'_\alpha$ and in the opposite $J_\alpha\gg J'_\alpha$ limit.  Although the effective Hamiltonian looks quite different in these two limits, in both cases it is straightforward (but tedious) to show that $E_0$ is minimized by a uniform flux configuration $\{\phi_p\} = \{1\}$ (or the flux configuration related by $\hat T$). Numerically, we have computed $E_0$ as a function of the ($\vec \Phi,\{\phi_p\}$) for a set of finite size systems with about $10^4$ sites and for various set of $J_\alpha$ and $J'_\alpha$. In all cases, we found that $E_0$ is minimized by the same uniform flux configuration. Thus, the lattice translational, point group, and spin symmetries~\cite{Baskaran07} are preserved. For a finite system with PBC, the energy difference between ground states with different global fluxes decays exponentially with the linear dimension of the system for gapped Majorana fermions. In the thermodynamic limit, for the Abelian CSL, the ground state degeneracy is fourfold coming from four possible values of the global fluxes, $\Phi_x=\pm 1$ and $\Phi_{y}=\pm 1$. However, for the non-Abelian CSL, the ground state is threefold degenerate because the projection operator annihilates one global flus ground state~\cite{comment_wave_function,comment_degeneracy}. We thus focus on the uniform flux (ground state) sector. The corresponding free Majorana Hamiltonian, ${\cal H}_{\textrm{UF}}$, is of the same form as Eq.~(\ref{eq_Ham_gauge}) with all $\hat U_{ij}$ and $\hat F_m$ replaced by 1.

${\cal H}_{\textrm{UF}}$ can easily be exactly diagonalized. Each site is labeled by a unit cell index $\br$ and a site index $\alpha=1,\cdots, 6$ within the unit cell as shown in Fig.~\ref{fig_triangle_honeycomb_lattice}(b) so
\begin{eqnarray}
\nonumber
&&{\cal H}_{\textrm{UF}}=i\sum_{\br}\Big[J'_x(c_{\br,4}c_{\br+\be_2,1}) +J_x(c_{\br,5}c_{\br,6} +c_{\br,3}c_{\br,2})\\\nonumber
&&\qquad\qquad~~+J'_y(c_{\br,2}c_{\br+\be_1,5})+J_y(c_{\br,1}c_{\br,3} +c_{\br,6}c_{\br,4})\\ 
&&\qquad\qquad~~+J'_z(c_{\br,3}c_{\br,6}) +J_z(c_{\br,1}c_{\br,2}+c_{\br,4}c_{\br,5})\Big],
\label{fermion}
\end{eqnarray} 
where $\be_{1,2}=\hat \bx/2\pm\hat \by \sqrt{3}/2$ are two unit vectors shown in Fig.~\ref{fig_triangle_honeycomb_lattice}(b).  We define a 6-component spinor field $\psi_{\alpha}(\bk) \equiv \sum_{\br} \te^{-i\bk\cdot\br}c_{\br,\alpha} / \sqrt{2N}$, where $N$ is the number of unit cells and $\bk$ lies in the first Brillouin zone.  Then,  
$ {\cal H}_{\textrm{UF}}=\sum_{\bk}\psi^\dag(\bk)  H(\bk) \psi(\bk)$ with $H(\bk)$ a $6\times 6$ hermitian matrix, and $H^\ast(\bk) = - H(-\bk)$. 
Since $c_{\br,\alpha}$ is a Majorana fermion, $\psi_{\alpha}(-\bk)=\psi_{\alpha}^\dag(\bk)$ which means that $\psi$ satisfies the usual 
anticommutation relations:  $\{\psi^\dag_{\alpha}(\bk),\psi_{\beta}(\bq)\} =\delta_{\bk\bq}\delta_{\alpha\beta}$.  Consequently, the problem is equivalent to a familiar free fermion problem, with the understanding that we must identify the states at $\bk$ and $-\bk$ to avoid double counting of states. There are six bands, $\epsilon_n(\bk)$, corresponding to six sites per unit cell. The spectrum is gapped except when $D(\bk)\equiv \Det[H(\bk)]=0$.

\begin{figure}[b]
\subfigure[]{
\includegraphics[scale=0.81]{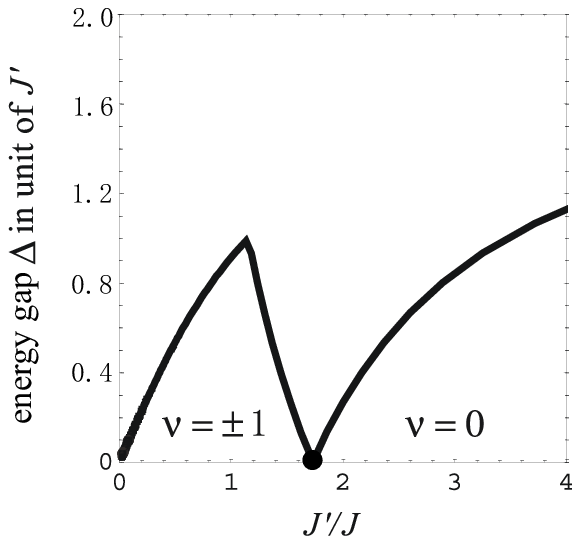}}
\subfigure[]{
\includegraphics[scale=0.158]{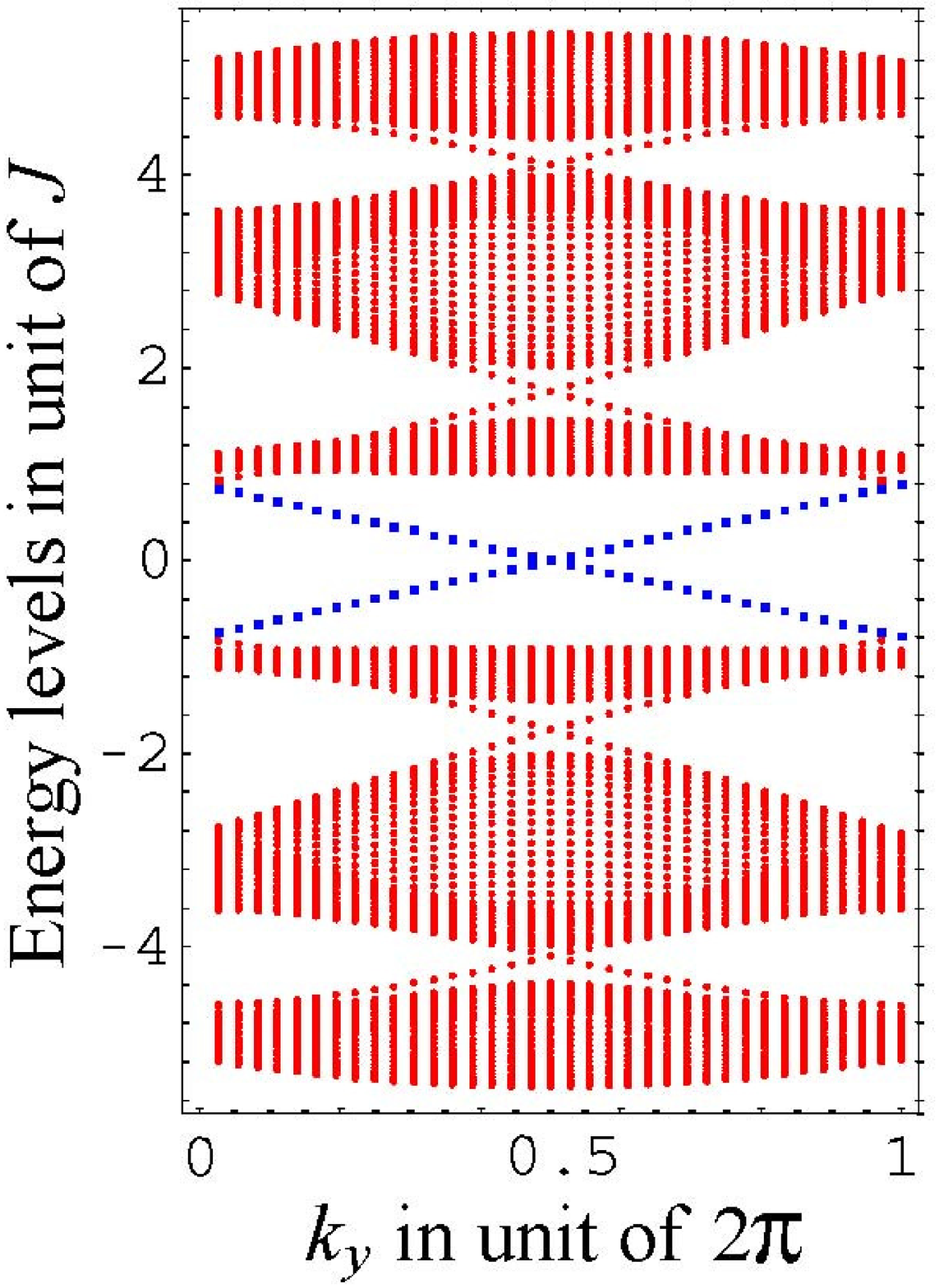}}
\caption{(a) Bulk energy gap, $\Delta$ 
as a function of $J^\prime/J$ for the symmetric case, $J_\alpha\equiv J$ and $J_\alpha^\prime\equiv J^\prime$.  
The quantum critical point is at $(J'/J)_\textrm{c}=\sqrt{3}$. For small $J$, the phase is a topologically trivial CSL with Abelian excitations while for large $J$, the phase is a topologically non-trivial CSL with vortex excitations obeying non-Abelian statistics.  The $J\to\infty$ 
spectrum is Dirac-like, with a mass-gap that vanishes as $J'^2/J$. 
There is a cusp at $J'/J=2\sqrt{3}/3$, where the $\bk$ with minimum energy jumps. (b) Energy levels calculated for 
$J=J^\prime$ on a cylinder with approximately $10^4$ sites and with OBC along $\hat x$-direction. Note that only half of the states are independent. There is one edge state (blue dotted line) crossing with zero energy, corresponding to Chern number $\nu=\pm 1$.} 
\label{newfig}
\end{figure}

{\it Chern number and edge states}:
To simplify the further discussion, we will henceforth consider the case $J_x=J_y=J_z\equiv J$ and $J_x^\prime=J_y^\prime=J_z^\prime\equiv J^\prime$.  We note that these conditions are implied if we want the Hamiltonian to respect the various mirror symmetries of the lattice pictured in Fig.~\ref{fig_triangle_honeycomb_lattice}(a).  
 
It is straightforward to show that $D(\bk) = 0$ only when $\bk = (\pi,-\pi/\sqrt{3})$ and $J^\prime= \sqrt{3}J$.  Thus, there is a quantum phase transition between two fully gapped phases which occurs at $J^\prime= \sqrt{3}J$, as  shown in Fig.~\ref{newfig}(a).  These two phases are topologically distinct as can be seen by directly computing the Chern number $\nu$:  for $J'> \sqrt{3}J$,  $\nu=0$, while for $J' < \sqrt{3}J$, $\nu=\pm 1$. That $\nu=\pm 1$ for $J'<\sqrt{3}J$ can be understood as follows: As $J\to \infty$, the spectrum of Eq.~(\ref{fermion}) to first order in $J'$ is gapless with Dirac-cones, as in graphene. Terms second order in $J'/J$ generate an effective Haldane mass term which breaks TRS~\cite{long}, gaps the Dirac cone, and gives $\nu=\pm 1$~\cite{Haldane88}. The same conclusion can be reached by computing the edge states of the system on a cylinder, since they reflect the topological character of the bulk state~\cite{Hatsugai93}. In the small $J$ (topologically trivial) phase, there is a gap about zero energy in the edge state spectrum, as well as the bulk.  However, in the large $J$ (topologically non-trivial) phase, there is an edge mode which disperses through zero energy.  This is illustrated in Fig. \ref{newfig}(b), in which the spectrum for $J^\prime=J$ of a relatively large cylinder (of order $10^4$ sites) is plotted as a function of $k_y$. We see that there is exactly one edge state which crosses the zero Fermi energy, corresponding to $\nu=\pm 1$. Thus, there exists a topological quantum phase transition at $(J'/J)_\textrm{c}=\sqrt{3}$, where the bulk gap vanishes and the edge states reconnect. [See Fig. \ref{newfig}(a).] 

{\it Non-Abelian anyons}:
The ground state with uniform $\{\phi_p\}=\{1\}$ is, by definition,  vortex-free. Vortex excitations are defined on plaquettes with $\phi_p=-1$. Since consistency requires $\prod_p \phi_p=1$, vortices can only be created in pairs. In all cases that we have studied, the vortex excitations always cost a non-vanishing energy, {\it i.e.} they are gapped.  It is presumably a peculiarity of the Kitaev model that the vortices have no dynamics, even when they interact with each other.  

It is to be expected that the vortex excitations in a state with odd Chern number obey non-Abelian braiding statistics~\cite{Kitaev06}, similar to the physics of $p+ip$ superconductors~\cite{Ivanov01} and the Pfaffian state of 5/2-FQHE~\cite{Moore_Read_91}. Specifically, in the odd Chern number phase, each vortex carries a Majorana zero energy mode. A pair of Majorana zero-energy modes on two well separated vortices constitute a single complex fermionic mode with zero energy. Taking into account the four possible global fluxes and the effects of the projection operator~\cite{comment_wave_function}, the ground state degeneracy approaches $2^{n+1}$ for a system of  $2n$ well separated vortices. We have numerically calculated the spectrum for various configurations of even numbers of vortices in a set of finite lattices with sizes about $10^4$ sites.  In all cases, we find the expected number of zero modes. Thus, we have confirmed that the ground state degeneracy for $2n$ vortices on a torus is $2^{n+1}$, independent of their positions or whether the vortices are associated with a triangular or a dodecagonal plaquette, so long as the vortices are separated by a distance large compared to the inverse energy gap. We interpret this correct degeneracy as confirming the non-Abelian statistics of the vortices.

In the small $J$ ($\nu=0$) phase, the vortices have Abelian braiding statistics.  When $J\ll J'$, by mapping the model to an effective $Z_2$ gauge theory on a ``dual'' honeycomb lattice with spontaneously broken TRS~\cite{long}, we find the vortex anyons on both triangle and dodecagon plaquettes are bosons. 
Interestingly, they are relative semions.  

{\it Discussion}:
As far as we know, the present model is the first exactly solvable model of a CSL~\cite{martin}.  The ground state spontaneously breaks TRS and is gapped, but does not break any spin rotational or lattice symmetries of the Hamiltonian. It is difficult to identify a state by what it does not do, which is a problem with spin liquids.  Presumably, a CSL will have signatures of TRS breaking such as a zero-field Kerr effect~\cite{Kapitulnik06} and thermal Hall conductance~\cite{Kane97}. 
More generally, the effective low energy theory of the non-Abelian phase is probably related to $SU(2)_2$ Chern-Simon theory~\cite{Fradkin98}. The problem remains, of course, where to find this state in real materials.

We thank T. Xiang, S.-C. Zhang, E.-A. Kim, X.-L. Qi, K. Shtengel, J. Vidal, E. Fradkin, S. B. Chung and Z.-H. Wang for helpful discussions. This work was supported, in part, by D.O.E. grant  DE-FG02-06ER46287 any by a Stanford Graduate Fellowship (H.Y.).

Note added: Some of the present result were anticipated in a remark made by Kitaev in Ref.~\cite{Kitaev06}.

\vspace{-0.4cm}

\begin{thebibliography}{30}
\bibitem[{com({\natexlab{b}})}]{comment_wave_function}
\bibinfo{note}{There are $2^{L_y-1}$ gauge-like choices $\{U_{ij}\}$ on
  $z'$-links consistent with the specified fluxes $(\vec \Phi,\{\phi_p\})$---here, $L_y$ is the width of the system along $\hat y$ direction. The
  eigenstate $\psi(\{U_{ij}\})$ of the corresponding quadratic Hamiltonian is {\it not} an eigenstate of $ \hat F_m$ or ${\cal H}$. (For OBC, it is an
  eigenstate of ${\cal H}$ since $\hat F_m$'s do not enter ${\cal H}$. It follows that the ground state with OBC has an additional $2^{L_y-1}$-fold degeneracy.) Nonetheless, the true eigenstate $\Psi(\vec \Phi,\{\phi_p\})$ can be projected out by applying $\hat P(\{F_m\})=\prod_m (1+F_m\hat F_m)/2$, i.e.,
  $\Psi(\vec \Phi,\{\phi_p\})=\hat P(\{F_m\}) \Psi(\{\hat U_{ij}\})$. Since
  $[\hat F_m, {\cal H}]=0$, the projection does not affect the eigenenergy.
 One can show that half of the eigenstates $\psi(\{U_{ij}\})$ in each flux sector
  $(\vec \Phi,\{\phi_p\})$ are annihilated by $\hat P(\{F_m\})$~\cite{long}.}
\bibitem[{com({\natexlab{c}})}]{comment_T_breaking}
\bibinfo{note}{The two-fold degeneracy was mentioned in ref.~\cite{Kitaev06}.}
\end{thebibliography}

\end{document}